\def\year{2023}\relax
\documentclass[hidelinks, letterpaper]{article} 
\usepackage{aaai22}  
\usepackage{times}  
\usepackage{helvet}  
\usepackage{courier}  
\usepackage[hyphens]{url}  
\usepackage{graphicx} 
\usepackage{natbib}  
\usepackage{caption} 
\DeclareCaptionStyle{ruled}{labelfont=normalfont,labelsep=colon,strut=off} 
\usepackage{algorithm}
\usepackage{algorithmic}

\usepackage{newfloat}
\usepackage{listings}
\floatstyle{ruled}
\newfloat{listing}{tb}{lst}{}
\floatname{listing}{Listing}
\title{A Contribution to the Defense of Liquid Democracy}
\author{
    Gregory Butterworth and Richard Booth\\       
}
\affiliations{
    \textsuperscript{}Cardiff University, School of Computer Science and Informatics\\    

    butterworthg2@cardiff.ac.uk, boothr2@cardiff.ac.uk
}

\usepackage{bibentry}

\usepackage{tikz}
\usepackage{multirow}

\begin{document}

\maketitle

\begin{abstract}
Liquid democracy is a hybrid direct-representative decision-making process that provides each voter with the option of either voting directly or to delegate their vote to another voter, i.e., to a representative of their choice. One of the proposed advantages of liquid democracy is that, in general, it is \emph{assumed} that voters will delegate their vote to others that are better informed, which leads to more informed and better decisions. 
Considering an audience from various knowledge domains, we provide an accessible high-level analysis of a prominent critique of liquid democracy by Caragiannis and Micha. Caragiannis and Micha’s critique contains three central topics: 1. Analysis using their $\alpha$-delegation model, which does not assume delegation to the more informed; 2. Novel delegation network structures where it is advantageous to delegate to the less informed rather than the more informed; and 3. Due to NP hardness, the implied impracticability of a social network obtaining an optimal delegation structure.
We show that in the real world, Caragiannis and Micha’s critique of liquid democracy has little or no relevance. Respectively, our critique is based on: 1. The identification of incorrect $\alpha$-delegation model assumptions; 2. A lack of novel delegation structures and their effect in a real-world implementation of liquid democracy, which would be guaranteed with constraints that sensibly distribute voting power; and 3. The irrelevance of an optimal delegation structure if the correct result is guaranteed regardless.
We conclude that Caragiannis and Micha’s critique has no significant negative relevance to the proposition of liquid democracy.
\end{abstract}

\section{Introduction}
The UK political system of \emph{representative} democracy remains essentially the same as it was when \emph{political representatives}, i.e., MP’s (Members of Parliament), had to travel to the House of Commons by horse. In the UK and many other parts of the world, the numerous opportunities for technology to enhance the representative democratic process has not gone unnoticed.

For example, \emph{liquid democracy} \cite{Behrens2017, Blum2016, Valsangiacomo2022, PaulinOTYLDR2020} provides an opportunity to utilize the Internet to enable mass democratic participation. Notably, for `voters' that do not have the time to participate, and/or know somebody that is better informed, participation is enhanced with the provision of an option to delegate their vote. It should be noted that one of the proposed and \emph{assumed advantages} of liquid democracy is that, in general, voters will delegate to others that are better informed, which leads to more informed and better decisions. Liquid democracy may be implemented via e.g., \emph{liquid feedback} \cite{Behrens2014}, which enables the electorate to create polls, deliberate, delegate their vote or vote directly, and then potentially have their elected representative vote the same way e.g., in Parliament. Therefore, liquid democracy provides an opportunity to introduce an element of \emph{direct democracy} into a \emph{representative democracy} system to form a modern \emph{direct-representative hybrid} democracy. Notable examples of political parties that implement liquid democracy include the Pirate Party in Germany, Flux in Australia, and the Net Party (Partido de la Red) in Argentina.

However, the acceptability and uptake of the liquid democracy inspired direct-representative hybrid democracy proposal is hampered by various theoretical apprehensions. Recent research has explored such apprehensions and in some cases offered potential solutions.  Apprehensions include, but are not limited to: the power of users that have amassed a large number of delegated votes \cite[e.g.][]{Blum2016, Kling2015}; the failure of assumed delegation to the more informed to collectively arrive at better decisions than direct or representative democracy \cite[e.g.][]{ Caragiannis2019, Kahng2018}; voting security \cite[e.g.][]{Anwarul2022, Nejadgholi2021}; the practicalities of mass deliberation \cite[e.g.][]{ Landemore2021}; etc. Therefore, the widespread adoption of liquid democracy is hampered until it can allay such concerns and can be shown to be the most obviously preferable alternative to the democratic system that it intends to replace.

From the above apprehensions, we focus on the subject of liquid democracy’s assumed delegation to the more informed to collectively arrive at a better decision than direct democracy. This subject was notably analyzed by Kahng, Mackenzie, and Procaccia \shortcite{Kahng2018, Kahng2021}.

Kahng et al., investigate if there are ‘\emph{delegation mechanisms that are guaranteed to yield more accurate decisions than direct voting?}’, and conclude negatively, but point out that a non-local, i.e., centralized mechanism, that imposes a cap on the voting power of any delegate, will adequately distribute power so that liquid democracy is at least as accurate as direct voting. Kahng et al. introduce their delegation mechanism, which includes the constant $\alpha$ to denote the minimum assumed increase in voter competence level when an agent delegates their vote to another agent. This recent work was then critically analyzed by Caragiannis and Micha.

Caragiannis and Micha critically analyze and expand on the work of Kahng et al. with their: ‘\emph{A Contribution to the Critique of Liquid Democracy}’ \shortcite{Caragiannis2019}, which is apprehensive of liquid democracy when compared to direct democracy. Their critique of liquid democracy focuses on three main areas: 1. Challenging liquid democracy’s assumption of delegation to the more informed, and a critique based on their proposed alternative $\alpha$-delegation model of vote delegation, where delegation confirms intuition; 2. Providing novel delegation network structures that benefit from delegation to the less informed; and 3. Suggesting due to the theoretical NP hardness of finding an optimal delegation network structure, that real-world implementations would also be flawed due to their assumed inability to also discover such an optimal delegation network structure.

We suggest that in the real world, Caragiannis and Micha’s critique of liquid democracy has little or no relevance. Respectively, our critique is based on: 1. The identification of incorrect $\alpha$-delegation model assumptions. Specifically, that intuition is not the dominant factor that determines a delegation choice. Rather, delegation is determined predominantly by trust. Most importantly, most delegation choices fall outside the scope of $\alpha$-delegation because they are predominantly of a preset nature and involve no issue knowledge or associated ‘intuition’, and are based on reputation and trust; 2. A lack of novel delegation structures in a real-world delegation network,  and the negation of any significant negative effect by implementing constraints that sensibly distribute voting power; and 3. The irrelevance of an optimal delegation structure if the correct result is guaranteed regardless.

\subsection{Contribution}

Caragiannis and Micha’s prominent ‘\emph{A Contribution to the Critique of Liquid Democracy}’\shortcite{Caragiannis2019}, is cited as a valid critique of liquid democracy. Our contribution is an accessible analysis that reveals that their critique has, arguably, little to no significance in the real world.

\subsection{Related Work}
Our work is a direct response to the Caragiannis and Micha paper and aims to counter all their conclusions that are negative towards liquid democracy. To the best of our knowledge, there is no other paper that attempts to make such a complete, accessible, and high-level counterargument to their conclusions. Therefore, our work takes on a more philosophical nature, and has a focus of relating theory to the real-world, with the aim of creating a fuller understanding of the relevance of Caragiannis and Micha’s conclusions, and the discussion of the proposition of liquid democracy in general. To obtain the widest possible audience we deliberately avoid any demanding rigorous mathematical analysis.

Our work is most closely related to the very recent, independently conducted, work of Halpern et al. \shortcite{Halpern2021}, which also aims to counteract Caragiannis and Micha’s negative perspective of liquid democracy. We consider our work to be complementary.
Halpern et al., although also taking a real-world perspective, take a mathematical focus on the potential for delegated voting power to be distributed to too small a number of voters. Their mathematical delegation model is focused on other areas than ours.

We discuss at a general high level, issues surrounding Caragiannis and Micha’s ODP (optimal delegation problem). At a more formal and mathematical level, the ODP topic has recently been investigated by Becker et al. \shortcite{Becker2021}, and Zhang and Grossi \shortcite{Zhang2021}.

\section{Preliminaries}

We consider elections with a {\em ground truth}, i.e., with two voting options, one of which is objectively correct. There is a social network consisting of a set $N$ of $n$ voters, which are nodes connected by directed edges $E$ on a graph $G$, stated as $G(N,E)$.
Voters are also called agents. Agents may vote directly or delegate their vote to a connected agent. Delegations may be transitive but do not form cycles. Agents that both receive delegated votes {\em and} vote directly are often referred to as being a \emph{guru} or \emph{expert}. A guru or expert that has acquired a large number of votes is often referred to as being a \emph{super-voter}. The election result is calculated using weighted majority, with each guru having a weight equal to one plus the total number of delegations it receives. Each agent $i$ has a competence level $p\textsubscript{$i$}$ that indicates their probability of voting correctly. It is assumed there is some fixed parameter $\alpha$ such that agent $i$ may only delegate to agent $j$ if $p\textsubscript{$i$} + \alpha < p\textsubscript{$j$}$. This assumption is questioned by Caragiannis and Micha, leading them to propose their refined model, i.e., {\em $\alpha$-delegation}, which is described in the next section.

We use colloquial concepts of \emph{reputation} and \emph{trust}. Based on an agent's observation of another agent's past behaviour or performance, {\em reputation} is defined as the subjective qualitative belief that the observing agent has regarding the observed agent. {\em Trust} is defined as a measure of how much an agent (trustor) is willing to rely on the actions of another agent (trustee), and the situation is directed to the future. The quality of the reputation that the trustor has of the trustee, determines the amount of trust that the trustor has in the trustee.

\section{The Real-World Relevance of the ‘\emph{Ineffectiveness of Local Mechanisms}’ and the Concept of $\alpha$-delegations}

Caragiannis and Micha \shortcite{Caragiannis2019} analyze the liquid democracy delegation model proposed by Kahng et al. \shortcite{Kahng2018}, which includes a constant parameter, \emph{$\alpha$}, so that delegations from an agent $i$ to a neighboring agent $j$ can only occur if $p\textsubscript{j}$ $>$ $p\textsubscript{i}$ + $\alpha$. Caragiannis and Micha object to liquid democracy’s assumed delegation to the more informed, i.e., to those ‘\emph{neighbors that have strictly higher competence level}’.

Caragiannis and Micha’s main objection to the assumed delegation to the more informed, is based around their suggestion that it is unlikely that agents with low competence levels, e.g., $p\textsubscript{$i$}=10\%$, would delegate to those with high competence levels, e.g., $p\textsubscript{$j$}$ $= 70\%$, \emph{and we assume vice versa}. They discount this possibility on the grounds that, because a competence level can be interpreted as an indication of an agent's confidence in their choice - ‘\emph{A competence level of 10\% for agent $i$ can be interpreted as a 90\% confidence}’ - it follows that agent \emph{$i$} would not delegate to agent \emph{$j$}, because agent \emph{$i$} would assume agent \emph{$j$} is ‘\emph{highly misinformed}’. Caragiannis and Micha then go on to refine the assumed liquid democracy delegation model by adding constraints that force delegations so that ‘\emph{a voter can choose to delegate to someone who confirms her intuition and looks more expert.}’, i.e., ‘\emph{We refine the sufficient condition for a delegation from agent $i$ to agent $j$ to be $p\textsubscript{j}$ $>$ $p\textsubscript{i}$ + $\alpha$ if $p\textsubscript{i}$ $\geq$ $\frac{1}{2}$ and $p\textsubscript{j}$ $<$ $p\textsubscript{i}$ - $\alpha$ if $p\textsubscript{i}$ $\leq$ $\frac{1}{2}$}’, and they then coin the term `\emph{$\alpha$-delegation}' to describe delegations that follow their adapted model.

In essence, Caragiannis and Micha’s adaptation of Kahng et al.’s model essentially forces \emph{correct} agents, i.e., $p\textsubscript{$i$}$ $>$ ${0.5}$, to delegate to \emph{more correct} agents, i.e., $p\textsubscript{$j$}$ $>$ $p\textsubscript{$i$}$ + $\alpha$, and forces \emph{incorrect} agents, i.e., $p\textsubscript{$i$}$ $<$ ${0.5}$, to delegate to \emph{more incorrect} agents, i.e., $p\textsubscript{$j$}$ $<$ $p\textsubscript{$i$}$ - $\alpha$. We differentiate and refer to these two forms as respectively \emph{positive-$\alpha$-delegation}, or \emph{+$\alpha$-delegation}, and \emph{negative-$\alpha$-delegation}, or \emph{-$\alpha$-delegation}. 

\subsection{Incorrect assumptions of the $\alpha$-delegation model}

We suggest that Caragiannis and Micha are incorrect when they assume that delegation agent choice is determined by confirmation of intuition. Instead, we suggest that trust is the dominant determinant on delegation agent choice. Therefore, the  $\alpha$-delegation model, which is based on delegation that confirms intuition, has little to no relevance to delegation within a real-world implementation of liquid democracy. 

Let us remind ourselves that in a scenario of direct democracy it is neither desirable nor practical to have the entire electorate deliberate on every issue. Liquid democracy’s raison d’être is it’s ability to circumvent this problem by offering a delegation option to those voters that do not have the time to participate in the democratic process. Therefore, by design, liquid democracy referendums will often have the vast majority of votes cast \emph{automatically} and therefore \emph{passively} via \emph{preset} delegation options. It is important to note that these passive automatic delegations will have been set in advance of future issues, and with no awareness of any of those issue’s specific details. Note that, automatic passive preset delegations are distinct from \emph{manual active dynamic} delegations made \emph{manually} during \emph{active} participation in the deliberation phase of an issue: consider an agent that during the process of deliberation \emph{dynamically} decides to either delegate their vote rather than vote directly, or decides to redelegate their vote to another delegate that they think is preferable to their current delegate.

Having supplied some context, our next point of analysis requires an estimate for the percentage of votes that are made via automatic preset delegations.

However, we are not aware of a resource that provides such data for a real-world large-scale, demographically representative, mature and stable liquid democracy implementation – in fact we are not aware of such an idealized real-world implementation of liquid democracy. Regardless, observing the relatively small scale German Pirate Party via analysis by Kling et al. \shortcite{Kling2015} and the German Pirate Party themselves \cite{PP2014}, we can surmise and extrapolate an approximation. Kling et al.’s extensive analysis of the German Pirate Party focuses on the power of super-voters, and as a result does include a suitable temporal analysis of the ratio of direct to delegated votes, or what percentage of the delegated votes are preset or active. But, Kling et al. do show that over time, users continue to make more delegations, and also that there is a reduction in the number of active users to approximately 10\% of the total number of registered users: users are labelled inactive after 180 days without login. Also, considering the section of the German Pirate Party's statistics  that are based on their ‘01/28/2014’ data \cite{PP2014}, their is a general trend of an increasing proportion of delegated votes to approximately 80\% of the total vote. It should be pointed out that the German Pirate Party data does not distinguish between the delegated votes being preset or active, but, considering that users would have to login to actively delegate their vote, and there is a reduction of active logins to approximately 10\% of all users, it seems reasonable to assume that the vast majority of delegated votes are preset delegated votes. Taken together, this seems to suggest there is a trend of increasing preset delegations, which would not be unexpected considering that this is exactly how liquid democracy voting behaviour is expected to evolve, i.e., over time, voters will increasingly identify other voters that they are willing to trust with their preset delegated vote.

We suggest that this trend should be extrapolated to the point that, in a stable implementation of liquid democracy, the number of delegations would on average make up approximately 90\% of the vote, and at least 95\% of these delegated votes would be preset. We also suggest this is a reasonable figure for a large-scale implementation of liquid democracy, because this would represent a realistic design target of 10\% of the agents being actively involved in the thorough deliberation of an issue, and/or voting directly. 

The German Pirate Party statistics, based on their ‘08/11/2012’ data, show that the ratio of direct to delegated votes varies widely, from approximately 1:4 to 4:1, or put another way, the percentage of direct and delegated votes both vary from approximately 20\% to 80\%,  This variation is expected because some issues are much more important or interesting than others.
Obviously, when an issue has minimal importance, the percentage of the agents being actively involved in deliberation and/or voting directly would be lower than a 10\% design target, and when an issue is very important with very wide-ranging effects, the percentage of the agents being actively involved in deliberation and/or voting directly would be much higher than 10\%. It is conceded that in cases where a large proportion of the agents are involved in the deliberation process, and there is a high proportion of direct voting, it seems probable that this scenario is condusive to more dynamic delegation. However, in general, we suggest that for a stable large-scale implementation of liquid democracy, an average of approximately 90\% of the delegated votes will be preset.

Having, hopefully, justified an approximate average automatic preset delegation rate of 90\%, we now make the key point that these preset delegated votes, which are made in advance of an issue with the aim of avoiding issue awareness and deliberation, fall outside of the $\alpha$-delegation model because these voters cannot be attributed to have any sense of issue intuition. In such cases of preset delegation, voters will have no voting intuition and will have solely made their delegation choices previously, based on reputation and trust. 

For the active delegating voters, delegation would more often than not, also be decided on reputation and trust, rather than exclusively via confirming intuition. Caragiannis and Micha dismiss the possibility of a very incorrect agent delegating to a very correct agent, because the very incorrect agent would consider the very correct agent to be ‘\emph{highly misinformed}’. Instead, they suggest delegation would be based on confirming intuition. Crucially, Caragiannis and Micha ignore or dismiss that the delegating influence of being ‘\emph{highly misinformed}’ could be overcome by trust. We suggest that a very incorrect agent could very feasibly delegate to a very correct agent if there was a strong enough trust relationship that overcame the influence of the appearance of being ‘\emph{highly misinformed}’. In such a case it is conceded that there is a requirement of a strong relationship of trust. For example, consider an issue with a ground truth that relies on the understanding of a complex mathematical issue, and three socially connected agents with an associated $p\textsubscript{i}$ measure: an undergraduate Computer Science student, with $p\textsubscript{i}=0.1$; a Mathematics PhD student, with $p\textsubscript{i}=0.3$; and a Mathematics Professor, with $p\textsubscript{i}=0.9$. Now, in this scenario, if the Mathematics PhD student decides to delegate their vote, to whom would they delegate? We suggest, that based on trust and reputation, the Mathematics PhD student would be more likely to delegate to the Mathematics Professor rather than the undergraduate Computer Science student. Also, we suggest that the undergraduate Computer Science student should also be aware of their own relative shortcomings and it would seem logical that if they decided to delegate their vote it would also be to the Mathematics Professor.

However, $\alpha$-delegation excludes the possibility of an incorrect agent ever delegating to a slightly more correct agent. But in these cases, delegation to the `\emph{slightly more correct}' would now appear as a delegation to the ‘\emph{slightly misinformed}’. In these cases, delegation to a more correct agent is much more likely because there is a significantly reduced requirement for the level of trust to have a dominant influence on delegation.

Considering such negative-$\alpha$-delegation and regardless of intuition, for any delegation from the incorrect to the more incorrect to occur, there must be an element of trust, because otherwise an agent would not delegate their vote to an agent that they do not trust. This poses the question, how did the agent that is more incorrect acquire this trust? It is conceded that in a real-world liquid democracy implementation that has just been created, the social network would not exclusively contain merit-based relationships of trust based on historical reputation. Agents that claim to be experts could be charlatans. Therefore, it is possible that trust could be assumed without any merit, which is conducive to an amount of negative-$\alpha$-delegation. However, over time, the liquid democracy social network will review past poll outcomes and the associated votes of the supposed experts. This will lead to the identification of those experts who can and cannot be trusted. Therefore, over time a social network will learn who are the true experts and who are the charlatans, and will correctly apportion trust to deserving ‘\emph{correct}’ agents. It follows that over time, the number of incorrect agents that are also trusted will be significantly reduced, which will significantly reduce the likelihood of negative-$\alpha$-delegation.

In summary, we suggest that trust is the dominant influence on delegation. Considering that 90\% of delegations are preset and outside the scope of $\alpha$-delegation, and that we suggest a similar percentage of the remaining 10\% of delegations would also be based on trust, then we approximate that 99\% of delegations are not applicable to $\alpha$-delegation. Therefore, any $\alpha$-delegation based critical analysis of liquid democracy is weakened by a similar amount, essentially rendering any $\alpha$-delegation based critique to insignificance.

\section{The Real-World Relevance of ‘\emph{The Curse of $\alpha$-Delegations}’}
We next evaluate Caragiannis and Micha's second objection to $\alpha$-delegation, which they consider to be significantly more important, i.e., ‘\emph{The Curse of $\alpha$-Delegations}’, where they present structures where ‘\emph{delegating to less-informed agents is highly beneficial}’.

\emph{‘Example 1}’, provides what is on the surface, a compelling critique that contradicts a promoted assumed advantage of liquid democracy delegative voting, i.e., that delegative voting involves delegating to more competent voters, which results in more informed decision making with a higher chance of ground truth discovery. Caragiannis and Micha show that structures exist where delegating to less-informed agents is highly beneficial. The structure example consists of an odd number of agents, split into a minimal majority forming a network star shape, with the central agent having a competence level slightly less than the leaf agents, and the remaining maximal minority of agents being isolated. They state this as follows:

 ‘\emph{Example 1. Consider a set $N$ of $n$ agents (with odd $n$), connected through a social network $G(N,E)$ as follows: $(n - 1)/2$ agents are isolated while the remaining ones form a star. The competence levels $p$ are $p\textsubscript{c}$ $=$ 1$-$2$\epsilon$ for the center and p\textsubscript{l} = 1 $-$ $\epsilon$ for each leaf agent, where $\epsilon$  $>$ 0 is a negligibly small constant. The isolated nodes have competence level $p\textsubscript{i} = 0$.}’.
 
Figure 1 illustrates such a structure with $\emph{N} = 7$. Note the dotted arrows show potential delegations. The associated node values represent competence values, $p\textsubscript{i}$, that predict the probability of the voter voting for the issue’s ground truth.

\begin{figure}[ht]
\centering
\begin{tikzpicture}
\node[label=left:{0\%},shape=circle,draw=black] (A) at (0,0) {1};
\node[label=left:{0\%},shape=circle,draw=black] (B) at (0,-1) {2};
\node[label=left:{0\%},shape=circle,draw=black] (C) at (0,-2) {3};
\node[label=right:{99\%},shape=circle,draw=black] (D) at (4,0) {4};
\node[label=left:{98\%},shape=circle,draw=black] (E) at (4,-1) {5};
\node[label=left:{99\%},shape=circle,draw=black] (F) at (3,-2) {6};
\node[label=right:{99\%},shape=circle,draw=black] (G) at (5,-2) {7};
\path [->](D) edge [dashed] node[left] {} (E);
\path [->](F) edge [dashed] node[right] {} (E);
\path [->](G) edge [dashed] node[left] {} (E);
\end{tikzpicture}
\caption{Beneficial less-informed delegation.}
\label{fig1}
\end{figure}

Caragiannis and Micha then consider the scenario in which \emph{only delegations to more informed agents} are allowed and observe that this structure provides zero possible options for delegation to the more informed. Therefore, all agents must vote directly. It follows that a correct result is only possible if all star agents vote correctly, which is increasingly unlikely and approaches zero as \emph{$N$} grows, i.e., more and more likely for a single star agent to vote incorrectly and tip the balance to an incorrect result. Caragiannis and Micha, observe that when the external nodes (4, 6, and 7, on the right in figure 1) delegate to the lower competence value central node (5), there is a higher probability of ground truth discovery, compared to when all these nodes vote directly. Therefore, ‘\emph{delegating to less-informed agents is highly beneficial}’. A stinging attack indeed.

However, stepping back into the real world, some key observations can be made.

Considering that there is only a problem if a poll result does not discover the ground truth, it is natural that analysis includes scenarios where the vote is in the balance, such as illustrated in figure 1.

Taking a general perspective of Caragiannis and Micha’s observation, their observation has an effect when the \emph{loss of accuracy} caused by agents \emph{delegating to a less informed} agent is \emph{less damaging} than the potential for a \emph{single direct voting agent to vote incorrectly} and swing the result to an \emph{incorrect} one, which is more likely as \emph{$N$} is increased. But how often would a real-world scenario relevant to their observation occur? 

Within a real-world implementation of liquid democracy, complete with sensible constraints, e.g. power distributed across numerous gurus, it is accepted that networks similar to the above structure could possibly exist, but only as a \emph{small outlier number of subnetworks} within a much \emph{larger network}. The rest of the network would consist of a much larger number of subnetworks, consisting of structures that are assumed to be typical of a liquid democracy implementation.

\begin{figure}[ht]
\centering
\begin{tikzpicture}
\node[label=left:{100\%},shape=circle,draw=black] (A) at (0,0) {1};
\node[label=left:{100\%},shape=circle,draw=black] (B) at (0,-1) {2};
\node[label=left:{100\%},shape=circle,draw=black] (C) at (0,-2) {3};
\node[label=right:{1\%},shape=circle,draw=black] (D) at (4,0) {4};
\node[label=left:{2\%},shape=circle,draw=black] (E) at (4,-1) {5};
\node[label=left:{1\%},shape=circle,draw=black] (F) at (3,-2) {6};
\node[label=right:{1\%},shape=circle,draw=black] (G) at (5,-2) {7};
\path [->](D) edge [dashed] node[left] {} (E);
\path [->](F) edge [dashed] node[right] {} (E);
\path [->](G) edge [dashed] node[left] {} (E);
\end{tikzpicture}
\caption{Beneficial less-informed delegation.}
\label{fig2}
\end{figure}

Interestingly, the structure of figure 1, but with \emph{inverted} $p\textsubscript{i}$ values, i.e., the isolated nodes having a competence level $p\textsubscript{i} = 100$ etc., as illustrated in figure 2, exposes a desirable property of this structure - consisting of a \emph{very competent slight minority} and a \emph{very incompetent slight majority} - which is that in the real-world, as ‘\emph{$n$}’ increases so does the likelihood of direct voting by external nodes, and therefore of obtaining a \emph{correct decision from a very incorrect majority}. If we assume that this type of subnetwork structure would appear, as an outlier, as often as the one cited by Caragiannis and Micha, then it follows that their effects would cancel each other out.

We next consider Caragiannis and Micha’s next example, ‘\emph{Example 2}’, as illustrated in figure 3. Note, that the dashed arrows show the options for a possible delegation.

\begin{figure}[ht]
\centering
\begin{tikzpicture}
\node[label=left:{100\%},shape=circle,draw=black] (A) at (0,0) {1};
\node[label=left:{50\%},shape=circle,draw=black] (B) at (2,0) {2};
\node[label=right:{70\%},shape=circle,draw=black] (C) at (4,0) {3};
\node[label=left:{60\%},shape=circle,draw=black] (D) at (2,-2) {4};
\node[label=right:{50\%},shape=circle,draw=black] (E) at (4,-2) {5};
\path [->](D) edge [dashed] node[left] {} (C);
\path [->](D) edge [dashed] node[left] {} (B);
\path [->](E) edge [dashed] node[right] {} (C);
\end{tikzpicture}
\caption{Social network and competence levels for `\emph{Example 2}'. Possible delegations in dashed arrows.}
\label{fig3}
\end{figure}
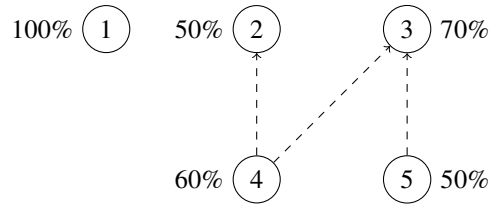

Caragiannis and Micha observe that considering all the possible delegation combinations, the possible competence levels for the network vary from 67.5\% to 85\%. We highlight the two most relevant combinations. Firstly, the highest probability occurs when agent 4 delegates to agent 2, and agent 5 delegates to agent 3, which highlights that ‘\emph{delegating to the most informed agent possible does not always equate to the optimal chance of the network discovering a ground truth}’. Secondly, the combination in which agents \emph{exclusively} delegate to agents with \emph{higher competence} levels, i.e., agents 4 and 5 both delegate to agent 3, which results in a \emph{lower network competence level} of 70\%.

Consider that an implication of Condorcet’s jury theorem \cite{Condorcet} is, assuming all voters have the same voting weight, and the average competence level of the voters is above 50\%, that as the number of voters is increased, a correct vote becomes a certainty. Now, considering the general case, and making the above assumptions, Caragiannis and Micha’s above observation has an effect when the \emph{loss of accuracy} caused by agents \emph{delegating to a less informed} agent is \emph{offset} by the \emph{increase in accuracy} caused by shifting the network structure closer to one that resembles a jury, i.e., equal voting weights and all competence levels hopefully above 50\%.

The relevance is that, in a real-world implementation of liquid democracy with sensible constraints, e.g., a minimum number of gurus with a voting weight limit, the network structure that Caragiannis and Micha describe would only exist at most as a small number of substructures in a much larger network. If the network structure was extended to represent a realistic number of voters, there would be no negative effect because the number of voters would become sufficient to ensure that a correct result is guaranteed via the Condorcet jury theorem.

In summary, it is our opinion that $\alpha$-delegation and the presence of rogue local subnetworks have no significant practical relevance that is terminal to the real-world proposition of liquid democracy.

\section{Optimal Delegation Problem (ODP)}
Caragiannis and Micha state the following: ‘\emph{We show that the optimal delegation problem (ODP) of maximizing the probability to find the ground truth by coordinating delegations is not only computationally hard but also hard to approximate with a substantial additive constant. The criticism by this complexity result should be clear: if it is hard to discover the ground truth in a coordinated way, why should the voters be expected to find it by acting independently?}’.

First, we make the general point that the proposition of liquid democracy does not fail because it cannot acquire an optimal chance of ground discovery via a perfect delegation structure, it fails if it cannot acquire a better chance of ground truth discovery than the incumbent system of democracy that it aims to replace, which is - if based on the motivations of the majority of real-world implementations of liquid democracy - representative democracy.

As mentioned in our ‘related work’ section, the ODP topic has recently been investigated in-depth by Becker \emph{et al}. \shortcite{Becker2021}, and Zhang and Grossi \shortcite{Zhang2021}. Becker  \emph{et al}., strengthen the existing hardness result, but counter this with an approximation result, which is dependent on assumptions relating to the competencies of the voters or the structure of the network. They also create simulations that show that on a large class of instances, liquid democracy outperforms direct democracy. Zhang and Grossi obtain an optimal group accuracy by central coordination, which uses ‘\emph{a variant of liquid democracy where proxies are weighted by a probability of delegation or, alternatively, by a share of the delegator’s voting weight}’, but achieve this by dropping Caragiannis and Micha’s network constraints of restricted connectivity.

Considering the above, and that a sensible implementation of liquid democracy could utilize a centralized system to limit guru power, it seems the ODP problem is overblown.

Trivially, we make the point that, if the correct result is essentially guaranteed regardless of optimization, then optimization is not an issue. Specifically, if the result of obtaining the ground truth is essentially guaranteed regardless of the voters finding an optimal delegation voting structure, i.e., via the Condorcet jury theorem, when there are enough agents and gurus with competence levels above 50\%, then the ODP is not an issue.  The ODP would only be relevant if an optimal delegation structure was the difference between the ground truth being discovered or not, but how often would this be the case if the liquid democracy implementation included quality deliberation and sensible voting mechanism constraints?

Also, a non-optimal delegation structure provides an opportunity for the improvement of subsequent delegation structures. If we consider that gurus that voted  \emph{incorrectly} will be identified, and then over time will become less likely to receive delegated votes, then a non-optimal delegation structure should contribute to the \emph{learning} of the social network and over time move it towards an optimal structure, or at least one that increasingly guarantees a correct result as the average competence levels of the gurus increase. 

Importantly, considering that all agents are subject to the real-world constraints of time, personal knowledge, and personal networking, it is not realistic to suggest or expect that all agents were ever going to independently coordinate their efforts to investigate every single possible delegation combination to arrive at an optimal chance of ground truth discovery. The central crux of Caragiannis and Micha’s criticism by their complexity result is not relevant because it is based on a proposed process that will never be a requirement of any liquid democracy implementation.

From a real-world perspective, it could be said that any final delegation structure is optimal considering the accuracy is subject to the above constraints. Arguably, a delegation structure that has been derived totally independently is theoretically optimal from the perspective of trust: trust being acquired via the transparency and the ownership of an individual’s decision to vote directly or delegate to an agent of their choice. The value of trust in any democratic system should not be ignored or underestimated.

Based on all of the above points, we consider Caragiannis and Micha’s ODP, and criticism by their complexity result to have no negative bearing on the proposition of liquid democracy.

\section{Discussion}
We make the general point that the negative effects that Caragiannis and Micha observe are associated with essentially small network structures that have power concentrated across a relatively small number of agents. Such network structures more closely resemble those found in representative democracies, e.g., UK government cabinet meetings \cite{Cabinet}, rather than those that would be found in large-scale liquid democracy implementations. We suggest that Caragiannis and Micha’s negative observations are arguably more applicable to representative democracy than liquid democracy.

Note that Caragiannis and Micha restrict their proofs in their \emph{Sections 3 and 4} to proxy voting, i.e., non-transitive voting, and in \emph{Theorem 1} highlight the effect of low values of $\alpha$ to reinforce the significance of their analysis. It is conceded that although a transitive delegation chain can be approximated by a series of proxy votes, it is not the same unless it accounts for the potential of multiple additions of $\alpha$ to be obtained via the delegation chain. Also, we point out that the value of $\alpha$ has a fundamental effect on the performance of liquid democracy, and unfair assumptions of the preponderance of low values of $\alpha$, provide critics of liquid democracy with an opportunity that in the real world we assume would not exist.

\section{Conclusion}
First, we conclude that in a real-world scenario of liquid democracy, the amount of $\alpha$-delegation is insignificant,  because trust rather than intuition is the dominant influence over delegation agent choice. Secondly, network structures that are observed to benefit from delegation to the less informed would be restricted to substructures with a minimal effect if sensible power distributing constraints are implemented to ensure the effect of the Condorcet jury theorem. Lastly the ODP problem is overblown, because a liquid democracy system utilizing such previous constraints to ensure the effect of the Condorcet jury theorem, would guarantee a correct result regardless of an optimal delegation structure.

In summary, Caragiannis and Micha’s critique is restricted to a narrow theoretical perspective that is not valid in a sensible real-world implementation of liquid democracy. We conclude that Caragiannis and Micha’s critique is not deleterious to the real-world proposition of liquid democracy.

\section{Future Work}

This work is intended to serve as an introduction and motivation for a future rigorous and comprehensive temporal analysis of the delegation process in general.


\bibliography {gb}

\end{document}